\documentclass[fleqn,11pt]{article}

\usepackage{amsmath,amsfonts,amssymb}

\usepackage{geometry}
\newtheorem{theorem}{Theorem}
\newtheorem{lemma}[theorem]{Lemma}

\newtheorem{proposition}{Proposition}

\newcommand{\R}{\mathbb{R}}

\newcommand{\T}{\mathbb{T}}
\newcommand{\ep}{\epsilon}

\newcommand{\Tr}{\rm{Tr}}

\newcommand{\cqfd}
{%
\mbox{}%
\nolinebreak%
\hfill%
\rule{2mm}{2mm}%
\newline
\newline
}

\begin{document}
\title  {ON LOW TEMPERATURE KINETIC THEORY;\\
SPIN DIFFUSION, BOSE EINSTEIN CONDENSATES, ANYONS}
\author{Leif Arkeryd\thanks{Mathematical Sciences, S-41296 Gothenburg, Sweden.}}
\date{}

\maketitle
\hspace{1cm}\\
\hspace{1cm}\\
{\bf Abstract}\\
\hspace{.1cm}\\
The paper considers some typical problems for non-linear kinetic models evolving through pair-collisions at temperatures not far from absolute zero, which illustrate specifically quantum behaviours. Based on these - mostly recent - examples, a number of differences between quantum and classical Boltzmann theory are then discussed in more general terms.\\
\\
\\
\hspace{1cm}\\
\setcounter{section}{-1}
\section{Introduction.}
An exciting area in present day physics is condensed matter physics - for its theory, its experiments, and its analogy models reaching from the standard
model of particle physics to cosmology and black holes (cf [V]). It also abounds with kinetic problems, quite interesting and challenging in themselves, and
shedding new light on classical kinetic theory. \\
\\
We first recall some basic facts about the classical Boltzmann case.
Consider a parcel of gas evolving as
\begin{eqnarray*}
\frac{dx}{dt}= v,\frac{dv}{dt}=F.
\end{eqnarray*}
That implies the evolution of the gas density $f$;
\begin{eqnarray*}
D_t f(x(t),v(t),t)=\frac{\partial f}{\partial t}+\frac{\partial f}{\partial x}\frac{dx}{dt}+\frac {\partial f}{\partial v}\frac{dv}{dt}=\frac{\partial f}{\partial t}+v\frac{\partial f}{\partial x}+F\frac {\partial f}{\partial v},
\end{eqnarray*}
which gives the Boltzmann equation when driven by a collision term $Q(f)$,
\begin{eqnarray*}
(D_t f(x(t),v(t),t)=)\frac{\partial f}{\partial t}+v\frac{\partial f}{\partial x}+F\frac {\partial f}{\partial v}=Q(f).
\end{eqnarray*}
\hspace{1cm}\\
\\
For two particles  having pre-collisional velocities $v,v_*$, denote the velocities after collision by $v',v'_*$, and write $f_*= f(v_*),\quad f'= f(v'),\quad f'_*= f(v'_*)$  . In this notation the classical Boltzmann collision operator becomes $Q(f)=\int B(f'f_*'-ff_*)dv_*d\omega$. The kernel $B$ is typically
of the type $b(\omega)|v-v_*|^s$ with $-3<s\leq 1$. The collision term is linear in the density of each of the two participating molecules, and all collisions respecting the conservation laws are participating. Mass and kinetic energy are each conserved, and the monotone in time entropy $\int dvdx f\log f(t)$ prevents strong concentrations to form.\\
\\
The quantum case is different from the classical one (cf [K], [S]). The rich phenomenology of low temperature gases corresponds to a similarly rich mathematical structure for its models, including the kinetic ones. Close to absolute zero, permitted energy levels are often discrete, implying fewer collisions. There may exist a condensate, where an excitation can interact only with those modes of its zero-point motion that will not give away energy. In the quantum regime, the thermal de Broglie wave length may become much larger than the typical
inter-particle distance, a situation  absent from classical kinetic theory. The scattering is wave-like, and
two-body quantum statistics gives an imprint on the individual collision process, e.g. the Pauli principle for fermions.
A quantum situation is usually not scale invariant; change the energy interval and the nature of participating particles may change, e.g. from atoms to phonons. When temperature passes a critical value, the description suggested by the old model may be wildly inadequate.\\
\\
In order to illustrate a number of differences from the classical case, this paper considers some typical non-linear quantum kinetic models evolving through pair collisions; \\
\hspace{1cm}\\
1) the Nordheim-Boltzmann model,\\
2) spin in the fermionic case,\\
3) low temperature bosons and a condensate,\\
4) a kinetic anyon equation.\\
\hspace{1cm}\\
These models have been obtained by phenomenological physics arguments from wave mechanics in the Heisenberg setting and from quantum field theory. A few studies of formal mathematical validation from interacting quantum particles can be found [ESY], [BCEP], [S2], and from weakly nonlinear wave equations with random initial data [LS].\\
\hspace{1cm}\\
Based on the examples 1-4, the final section will elaborate on some \\
5) differences between quantum and classical Boltzmann theory.\\
\hspace{1cm}\\
{\bf Acknowlegement.} The author wishes to warmly thank H. Spohn for valuable comments on an earlier version and  for providing the Fermi-Hubbard section 2c.\\
\\
\section{The Nordheim-Boltzmann (NB) model.}
In 1928 Nordheim [N] introduced the first quantum kinetic equation for bosons and fermions, followed a year later by Peierls' study of quantized waves (phonons) [P]. Uhling and Uhlenbeck in their 1933 paper [UU] discuss quantum corrections to the transport coefficients of real gases that must be present at low temperatures.
In these equations the
collision are rare enough that the motion of particles between collisions is classical,
and with it the
transport left hand side of these evolution equations. Only the collisional right hand side is directly influenced by the quantum effects. The term quantum Boltzmann equation usually refers to such equations, even though true quantum kinetic models are sometimes considered. For a more detailed discussion of such matters see [LS]. The boson case is of particular interest
in a neighbourhood of and below the transition temperature $T_c$, where a condensate first appears.
\\
\\
With the velocities $(v,v_*)$ before and $(v',v'_*)$ after a collision related by $v'= v-n[(v-v_*)\cdot n],\quad v'_*=v_*+n[(v-v_*)\cdot n]$,  we write $f_*=f(v_*), f'=f(v'), f'_*=f(v'_*)$. The collision operator for the NB equation is
\begin{eqnarray}
\begin{split}
{Q}_{NUU}(f)(p)= \int _{\R ^3\times \R ^3\times \R ^3}B\delta (p+p_*- (p'+p'_*))\delta (E(p)+E(p_*)- (E(p')+E(p'_*)))\\
\Big( f'f'_*(1+\epsilon f)(1+\epsilon f_*)-ff_*(1+\epsilon f')(1+\epsilon f'_*)\Big) dp_*dp'dp'_*,
\end{split}
\end{eqnarray}
where $B$ is a collision kernel, $E(p)=|p|^2$, $\epsilon= 1$ for bosons, $\epsilon=-1$ for fermions, and $\epsilon=0$ in the classical case. The quartic terms vanishes in all three cases. Validation for the fermion and boson cases is studied in [BCEP].\\
For $\epsilon=\pm 1$ the entropy is
\begin{eqnarray*}
\int\Big( f\log f-\frac{1}{\epsilon}(1+\epsilon f)\log (1+\epsilon f)\Big)dp.
\end{eqnarray*}
Also this entropy is monotone but, contrary to the classical case, gives no control of concentrations. In equilibrium $Q$ is zero, and multiplication by $\log\frac{f}{1+\epsilon f}$ and integration gives
\begin{eqnarray*}
0=\int Q(f) \log \frac{f}{1+\epsilon f}dp= \int_{\R ^3\times \R ^3\times \R ^3}B\delta (p+p_*-( p'+p'_*))\delta (E(p)+E(p_*)-( E(p')+E(p'_*)))\\
(1+\epsilon f)(1+\epsilon f_*)(1+\epsilon f')(1+\epsilon f'_*)\Big(\frac{f'}{1+\epsilon f'}\frac{f'_*}{1+\epsilon f'_*}-\frac{f}{1+\epsilon f}\frac{f_*}{1+\epsilon f_*}\Big)\\
\Big(\log\frac{f}{1+\epsilon f}\log\frac{f_*}{1+\epsilon f_*}-\log\frac{f'}{1+\epsilon f'}\log\frac{f'_*}{1+\epsilon f'_*}\Big).
\end{eqnarray*}
The two last factors in the integral have the same sign, which implies
\begin{eqnarray*}
\frac{f'}{1+\epsilon f'}\frac{f'_*}{1+\epsilon f'_*}\equiv \frac{f}{1+\epsilon f}\frac{f_*}{1+\epsilon f_*}.
\end{eqnarray*}
As in the classical case, we conclude that $\frac{f}{1+\epsilon f}=M$, a Maxwellian,
hence $f=\frac{M}{1-\epsilon M}$, which for $\epsilon = \pm 1$ is the {\it Planckian} equilibrium function.\\
There is a fairly large number of rigorous, low temperature results for the NB equation. Let us just mention a few. For fermions $(\epsilon=-1)$ the concentrations are bounded by one because of the factor $(1-f)$, which preserves positivity together with f. This is stronger than the entropy controlled concentrations from the classical case, and has been used to obtain general weak existence results for the fermion case in an $L^1\cap L^{\infty}$-setting, J. Dolbeault [D] and P.L. Lions [PLL]. In this fermion case there are also particular results on a weak type convergence to equilibrium [D], and in the space-homogeneous case also strong convergence to equilibrium [LW]. \\
Existence is more intricate in the boson case, lacking both direct and entropy based concentration bounds. So far there are only partial results - for the space-homogeneous, isotropic setting ([Lu1], [Lu2], [EMV] and others), and for particular space-dependent settings close to equilibrium ([R]). There are also a few results on convergence to equilibrium ([Lu3], [R]), and on oscillations, concentrations, condensations, and blow-up in finite time for the space-homogeneous, isotropic setting ([Lu3-4], [EV1-2]). \\
\hspace{1cm}\\
Results on existence, uniqueness and asymptotic behaviour have also been obtained ([EM]) for the related Boltzmann-Compton equation
\begin{eqnarray}
k^2\frac{\partial f}{\partial t} = \int_0^{\infty}(f'(1+f)B(k',k;\theta)-f(1+f')B(k,k';\theta))dk'
\end{eqnarray}
with $f$ photon density, $k$ energy, $\theta$ temperature, and with the detailed balance law $e^{k/\theta}B(k',k;\theta)=e^{k'/\theta}B(k,k';\theta)$.
This equation models a space-homogeneous isotropic photon gas in interaction with a low temperature gas of electrons in  equilibrium having Maxwellian distribution of velocities.\\
\\
\section {Some spin cases for Fermions}
\hspace{1cm}\\
{\bf 2 a) Spin in the fermionic case.}\\
\\
The experimental study of spin polarized neutral gases at low temperatures and their kinetic modelling is well established in physics, an early mathematical physics text in the area being [S]. The first experiments concerned very dilute solutions of $^3He$ in superfluid $^4He$ with - in comparison with classical Boltzmann gases - interesting new
properties such as spin waves (see [NTLCL]). The experimentalists later turned to laser-trapped low-temperature gases (see [JR]).
The mathematical study of these models, however, is less advanced.\\
To add spin to a Boltzmann equation for fermions, the density $f$ can be replaced by by a $2\times2$ hermitean matrix valued function. To discuss this, we first recall some properties of the Pauli spin matrices
\begin{eqnarray*}
\sigma_1=\left(\begin{array}{ll}0 & 1\\1 & 0\end{array}\right),\quad \quad
\sigma_2=\left(\begin{array}{ll}0 & i\\-i & 0\end{array}\right),\quad \quad
\sigma_3=\left(\begin{array}{ll}1 & 0\\0 & -1\end{array}\right).
\end{eqnarray*}
With $[\sigma_i,\sigma_j]$ denoting the commutator $\sigma_i\sigma_j-\sigma_j\sigma_i$, the Pauli matrices satisfy
\begin{eqnarray*}
\quad\quad
[\sigma_1,\sigma_2]=2i\sigma_3,\quad
[\sigma_2,\sigma_3]=2i\sigma_1,\quad
[\sigma_3,\sigma_1]=2i\sigma_2,\quad
{\rm and}\quad  [\sigma_i,\sigma_i]=0\quad{\rm for}\quad i=1,2,3,
\end{eqnarray*}
With $\sigma=(\sigma_1,\sigma_2,\sigma_3)$ the Pauli spin matrix vector,
this  is equivalent to $\sigma \times \sigma =2i\sigma$.
Let ${\mathcal M}_2(\mathbb C)$ denote the space of $2\times2$ complex matrices, and ${\mathcal H}_2(\mathbb C)$ the subspace of hermitean matrices. ${\mathcal H}_2(\mathbb C)$ is linearly isomorphic to $\mathbb R ^4$, if we use the decomposition $\rho=A_cI+A_s\cdot \sigma$ and identify $\rho\in{\mathcal H}_2(\mathbb C)$ with $(A_c,A_s)\in\mathbb R^4$.\\
A {\it dilute spin polarized gas} with spin $\frac{1}{2}$, can be modelled by a distribution function matrix $\rho\in{\mathcal H}_2(\mathbb C)$  which is  the Wigner transform of the one-atom spin-wave density operator for the system.
The domain of $\rho(t,x,p)$ is taken as positive time $t$, $p\in \mathbb R^3$, and for simplicity here with position-space $x$ periodic in 3d with period one. We shall in sections 2a and 2b focus on the following model for the kinetic evolution of $\rho$,
\setcounter{equation}{0}
\begin{eqnarray}
D\rho:=\frac{\partial}{\partial t} \rho + p\cdot \bigtriangledown_x\rho=Q(\rho)
\end{eqnarray}
with (in the Born approximation) the collision integral of [JM]
\begin{eqnarray}
Q(\rho)=
\int dp_2 dp'_{1}dp'_{2}B\delta(p_1+p_2-p'_{1}-p'_{2})
\delta(p_1^2+p^2_2-p'^{2}_{1}-p'^2_{2})\nonumber\\
\Big( \{ [\rho_{1'},\tilde{\rho}_1]_+\Tr(\tilde{\rho}_2\rho_{2'})
-[\tilde{\rho}_{1'},\rho_1]_+\Tr(\rho_2\tilde{\rho}_{2'}) \}\nonumber
-\{[\rho_{1'}\tilde{\rho}_2\rho_{2'},\tilde{\rho}_1]_+
-[\tilde{\rho}_{1'}\rho_2\tilde{\rho}_{2'},\rho_1]_+\}\Big).
\end{eqnarray}
Here $\tilde{\rho}=I-\rho$, and $[.,.]_+$ denotes an anti-commutator.
The number density of particles of any spin component, is given by $f:= Tr(\rho(t,x,p))$, and the magnetization of particles is given by the vector $\bar{\sigma}(t,x,p):= Tr(\sigma\rho(t,x,p))$, which implies $\rho=\frac{1}{2}(fI+\bar{\sigma}\cdot\sigma)$.
The resulting equations for $f$ and $\bar{\sigma}$ are
\begin{eqnarray}
Df= Q_n(f,\bar{\sigma}),  \\
D\bar{\sigma}=Q_m(f,\bar{\sigma}),
\end{eqnarray}
where
\begin{eqnarray*}
Q_n(f,\bar{\sigma})= \frac{1}{2}\int dp_2dp'_1dp'_2B\delta(p_1+p_2-p'_1-p'_2)\delta(p^2_1+p^2_2-p'^2_1-p'^2_2)\\
\Big(([f_{1'}-\frac{1}{2}(f_1f_{1'}+\bar{\sigma}_1\cdot\bar{\sigma}_{1'})][f_{2'}
-\frac{1}{2}(f_2f_{2'}+\bar{\sigma}_2\cdot\bar{\sigma}_{2'})]
-[f_1-\frac{1}{2}(f_1f_{1'}+\bar{\sigma}_1\cdot\bar{\sigma}_{1'})]
[f_2-\frac{1}{2}(f_2f_{2'}+\bar{\sigma}_2\cdot\bar{\sigma}_{2'})])\\
-([\bar{\sigma}_{1'}-\frac{1}{2}(f_{1'}\bar{\sigma}_1+f_1\bar{\sigma}_{1'})]
\cdot[\bar{\sigma}_{2'}
-\frac{1}{2}(f_{2'}\bar{\sigma}_2+f_2\bar{\sigma}_{2'})]-[\bar{\sigma}_1-
\frac{1}{2}(f_{1'}\bar{\sigma}_1+f_1\bar{\sigma}_{1'})][\bar{\sigma}_2-
\frac{1}{2}(f_{2'}\bar{\sigma}_2+f_2\bar{\sigma}_{2'})])\Big),\\
Q_m(f,\bar{\sigma})=\frac{1}{2}\int dp_2dp'_1dp'_2B\delta(p_1+p_2-p'_1-p'_2)\delta(p^2_1+p^2_2-p'^2_1-p'^2_2)\\
\Big(([f_{2}-\frac{1}{2}(f_2f_{2'}+\bar{\sigma}_2\cdot\bar{\sigma}_{2'})]
[\bar{\sigma}_{1}-\frac{1}{2}(f_{1'}\bar{\sigma}_1+f_1\bar{\sigma}_{1'})]-
[f_{2'}-\frac{1}{2}(f_2f_{2'}+\bar{\sigma}_2\cdot\bar{\sigma}_{2'})]
[\bar{\sigma}_{1'}-\frac{1}{2}(f_{1'}\bar{\sigma}_1+f_1\bar{\sigma}_{1'})])\\
-([f_{1'}-\frac{1}{2}(f_1f_{1'}+\bar{\sigma}_1\cdot\bar{\sigma}_{1'})]
[\bar{\sigma}_{2'}-\frac{1}{2}(f_{2'}\bar{\sigma}_2+f_2\bar{\sigma}_{2'})]
-[f_{1}-\frac{1}{2}(f_1f_{1'}+\bar{\sigma}_1\cdot\bar{\sigma}_{1'})]
[\bar{\sigma}_{2}-\frac{1}{2}(f_{2'}\bar{\sigma}_2+f_2\bar{\sigma}_{2'})]
)\Big).
\end{eqnarray*}
The collision term $Q_n$ does not change the number density ($\int Q_ndp =0$), the linear momentum density
($\int pQ_n dp=0$),
and the energy density ($\int p^2Q_ndp=0$),
and the collision term $Q_m$ does not change the magnetization density ($\int Q_mdp=0$).\\
\\
{\bf 2 b) Existence results.}\\
\\
Consider the initial value problem for the equations (2.2-3) in a periodic box. The initial value $(f_0,\bar{\sigma}_0)\in L^{\infty}$ is assumed to satisfy $0\leq f_0\leq 2$, $\min((2-f_0)^2,f_0^2)\geq \bar{\sigma}_0\cdot\bar{\sigma}_0$. We first treat the case with a truncation $B_j$ in the domain of integration for $Q$, where $B_j$, assumed bounded, is the restriction of $B$ to the set $p_1^2+p_2^2\leq j^2$.\\
Set
\begin{eqnarray*}
F(t,x,p)=f(t,x,p) \hspace{.3cm} {\rm for} \hspace{.5cm} 0\leq f\leq 2, \hspace{.5cm} =0 \hspace{.3cm} {\rm for} \hspace{.5cm} f<0, \hspace{.3cm}
=2 \hspace{.5cm} {\rm for} \hspace{.3cm} f>2,\\
\Sigma(t,x,p)=
\bar{\sigma}(t,x,p)\hspace{.5cm} {\rm when} \hspace{.3cm}  \min(F^2,(2-F)^2)\geq \bar{\sigma}\cdot\bar{\sigma},\hspace{3.7cm}\\ {\rm else} \hspace{.3cm}\Sigma(t,x,p)=
\frac{\min(F,2-F)\bar{\sigma}}{\sqrt{\bar{\sigma}\cdot\bar{\sigma}}}(t,x,p).
\hspace{4cm}
\end{eqnarray*}
The system $Df=Q_n(F,\Sigma), D\bar{\sigma}=Q_m(F,\Sigma)$ with the initial value $(f_0,\bar{\sigma}_0)$ can, under the truncation $B_j$, be solved by a contraction argument.
\begin{lemma}
{\rm [A2]} The initial value problem with initial values $(f_0,\sigma_0)$ for the truncated problem $ Df= Q_n(F,{\Sigma}), D\bar{\sigma}=Q_m(F,{\Sigma})$ with truncation $B_j$, has a unique hermitean local solution
in $L^{\infty}$.
\end{lemma}
It remains to prove that $F=f$ and $\Sigma=\bar{\sigma}$ and extending to a global result.
We first consider the special  initial data, such that for some $\eta_j>0$ and for all $|p|\leq j$, uniformly in $x$
\begin{eqnarray*}
0<\eta_j\leq f_0\leq2-\eta_j,\quad\bar{\sigma}^2_0+\eta_j^2\leq\min(f_0^2,(2-f_0)^2).
\end{eqnarray*}
Then $F=f$ and $\Sigma=\bar{\sigma}$ holds by continuity on a (short, $\eta$- and $j$-dependent) time interval
$0\leq t<t_j$, using the bounds for the contracted solution. We give a proof similar to [D] for general $t$, when there is spin only in the $\sigma_3$ direction, and refer to the paper [A2] for the general case.  With $\bar{\sigma}= (0,0, s)$, we can replace $\bar{\sigma}$ with $s$  in $Q_n$ and $Q_m$, and the equation for $\bar{\sigma}$ by the corresponding one for $s$. Set $c_1=18\max_{p_1} \int dp_2 dp'_1 dp'_2 B_j\delta(p_1+p_2-p'_1-p'_2) \delta(p^2_1+p^2_2-p'^2_1-p'^2_2)$. It holds
\begin{eqnarray}
-c_1(f^2_1-s^2_1)\leq \int dp_2 dp'_1 dp'_2B_j \delta(p_1+p_2-p'_1-p'_2) \delta(p^2_1+p^2_2-p'^2_1-p'^2_2)\nonumber\\
\frac{1}{2}\Big(-(f_1-s_1)\big([(f_1+s_1)(1-\frac{1}{2}(f_{1'}
+s_{1'}))]
[(f_{2}-s_2)(1-\frac{1}{2}(f_{2'}-s_{2'}))]\big)\nonumber\\
-(f_1+s_1)\big([(f_1-s_1)(1-\frac{1}{2}(f_{1'}
-s_{1'}))]
[(f_{2}+s_2)(1-\frac{1}{2}(f_{2'}+s_{2'}))]\big)\Big)\nonumber\\
+\frac{1}{4}\Big(-(f_1-s_1)\big([(f_1+s_1)(1-\frac{1}{2}(f_{1'}
+s_{1'}))]
[(f_{2}+s_2)(1-\frac{1}{2}(f_{2'}+s_{2'}))]\big)\nonumber\\
-(f_1+s_1)\big([(f_1-s_1);(1-\frac{1}{2}(f_{1'}-s_{1'}))]
[(f_{2}-s_2)(1-\frac{1}{2}(f_{2'}-s_{2'}))]\big)\Big)\nonumber\\
\leq (f_1-s_1)(Q_n+ Q_m)(f,s)+(f_1+s_1)(Q_n- Q_m)(f,s)=D(f^2_1-s^2_1).
\end{eqnarray}
This implies
\begin{eqnarray}
0<(f_0^2- s_0^2)(x,p)e^{-c_1t}\leq (f^2- s^2)^{\#}(t,x,p),
\end{eqnarray}
where $(f,s)$ is the contraction solution with initial value $(f_0,\bar{\sigma}_0)$ . By continuity $f\pm s>0$ for $0\leq t\leq t_j$. Analogously, starting from the equation for $(1-\frac{1}{2}(f_1\pm\sigma_1))$ instead of $f_1\pm\sigma_1$, we get by uniqueness the same solution, together with the estimate
\begin{eqnarray}
0<((2-f_0)^2- s_0^2)(x,p)e^{-c_1t}\leq ((2-f)^2- s^2)^{\#}(t,x,p),
\end{eqnarray}
which by continuity holds for $0\leq t\leq t_j$ together with $2-f\pm s>0$ for $0\leq t \leq t_j$. And so uniformly in $(x,p)$, $|s|<\min(f,2-f)\leq 1$ for $0\leq t\leq t_j$. By iteration, existence and uniqueness follow for $t>0$. We conclude that $\rho=\frac{1}{2}(fI+\bar{\sigma}\sigma)$, solves the truncated initial value problem for (2.1) globally in time. Approximating by the above type of uniformly positive initial values, global existence in the $B_j$-case follows for arbitrary initial values with $0\leq f_0\leq 2$, $\min((2-f_0)^2,f_0^2)\geq \bar{\sigma}_0\cdot\bar{\sigma}_0$.\\
\\
We now switch to the usual velocity-angular variables $dp_2d\omega$ with $\omega=(\theta,\varphi)$, and assume that the collision kernel $B(z,\omega)$ satisfies i) $0\leq B\in L^1(B_R\times\mathcal{S}^2):=L^1(\{z\in\mathbb{R}^3\;|z|\leq R\}\times\mathcal{S}^2)$ for $R>0$, ii) $B(z,\omega)=B(|z|,|(z,\omega)|)$,
iii) $(1+|z|^2)^{-1}\int_{z+B_R}\int_{\mathcal{S}^2}B(z,\omega)d\omega\rightarrow 0,$ when $|z|\rightarrow \infty,$ $R\in (0,\infty)$. Then the earlier truncation in $B_j$ can be removed using a variant of the limit procedure from [PLL] for the corresponding solutions $\rho_j$ above, when $j\rightarrow \infty$.
\begin{theorem}{\rm [A2]} Suppose that $(f_0,\bar{\sigma}_0)\in L^{\infty}\cap L^1([0,1]^3\times \mathbb {R}^3)$, and that
$0\leq f_0\leq 2$ and $\bar{\sigma}_0^2\leq \min(f_0^2,(2-f_0)^2)$. Then the system (2.2-3) with initial value $(f_0,\bar{\sigma}_0)$ , has a bounded integrable, periodic, weak solution for $t>0$ with $0< f< 2$ and $\bar{\sigma}^2< \min(f^2,(2-f)^2)$ .
\end{theorem}
For details see [A2]. \\
\\
Other problems of considerable physical interest in this context concern (cf [JM]) the relaxation times for spin-diffusion, the time asymptotic behaviour in general, and the influence of more involved collision and transport terms in (2.2-3) such as a physicists' version of the problem,
\begin{eqnarray*}
\frac{\partial f}{\partial t}+\bigtriangledown _p \epsilon_p\cdot \bigtriangledown_r f_p-\bigtriangledown_r\epsilon_p\bigtriangledown_p \cdot  f_p+\sum_{i=xyz}\Big[\frac{\partial h_p}{\partial p_i}\cdot \frac{\partial \bar{\sigma_p}}{\partial r_i} - \frac{\partial h_p}{\partial r_i}\cdot \frac{\partial \bar{\sigma_p}}{\partial p_i}\Big] =Q_n        \\
\frac{\partial \bar{\sigma_p}}{\partial t}+\sum_i\Big[ \frac{\partial \epsilon _p}{\partial p_i}\frac{\partial \bar{\sigma}_p}{\partial r_i}-
 \frac{\partial \epsilon _p}{\partial r_i}\frac{\partial \bar{\sigma}_p}{\partial p_i}+ \frac{\partial f _p}{\partial r_i}\frac{\partial h_p}{\partial p_i}- \frac{\partial f_p}{\partial p_i}\frac{\partial h_p}{\partial r_i}\Big]-2(h_p\times \bar{\sigma_p})=Q_m.
\end{eqnarray*}
Here
\begin{eqnarray*}
\epsilon_p=\frac{p^2}{2m}+\int dp'\{V(0)-\frac{1}{2}V(|p-p'|)\}f_{p'}(t,r,p')    \\
h_p=-\frac{1}{2}(\gamma B-\int dp' V(p-p')\bar{\sigma}_{p'}(t,r,p')),
\end{eqnarray*}
$V$ is the inter-particle potential, {B} an external magnetic field, and $\gamma$ is the gyromagnetic ratio.\\
\\
\\
In {\bf spintronics} for semiconductor hetero-structures a related {\it linear} Boltzmann equation is considered,
\begin{eqnarray}
\frac{\partial}{\partial t} \rho + v\cdot \bigtriangledown_x\rho
+E\cdot  \bigtriangledown_v\rho =Q(\rho) + Q_{SO}(\rho)+Q_{SF}(\rho).
\end{eqnarray}
Here $E$ is an electric field and $Q$ is the collision operator for collisions without spin-reversal, in the linear BGK approximation
\begin{eqnarray*}
 \int _{\mathbb{R}^3} \alpha(v,v')(M(v)\rho(v')-M(v')\rho(v))dv',
\end{eqnarray*}
$M$ denoting a normalized Maxwellian. The spin-orbit coupling generates an effective field $\Omega$ making the spins precess. The corresponding spin-orbit interaction term $Q_{SO}(\rho)$ is given by $\frac{i}{2}[\Omega\cdot \sigma, \rho]$. Finally $Q_{SF}(\rho)$ is a spin-flip collision operator, in relaxation time approximation given by
\begin{eqnarray*}
Q_{SF}(\rho)= \frac{{\rm tr} \rho I_2-2 \rho}{\tau_{sf}},
\end{eqnarray*}
with $\tau _{sf}>0$ the spin relaxation time.\\
Mathematical properties such as existence, uniqueness, and asymptotic behaviour, have been studied in particular by the French group around Ben Abdallah with coworkers and students (see [EH1], [EH2] and references therein).\\
\\
{\bf 2 c) Weakly interacting Fermi-Hubbard model}\\
\\
The Hubbard model describes the dynamics of electrons moving in a
fixed periodic lattice potential, which is taken into account through a tight binding approximation. Hence the
electrons hop on the lattice $\mathbb{Z}^3$. The momentum space
is then the 3-torus $\mathbb{T}^3$. The quadratic dispersion, $k^2$,
of the previous example is replaced by a smooth function $\omega:
\mathbb{T}^3 \to \mathbb{R}$. This is a common feature of all lattice models
and forces another level of complication in the analysis. Since the electrons carry a spin,
in the spatially homogeneous case the one-particle Wigner function, $W$, is defined on $\mathbb{T}^3$
and takes values in $\mathcal{H}_2(\mathbb{C})$. The Fermi property is codified as
$0 \leq W(k) \leq 1$. For weak interactions  the Wigner
matrix-valued function is governed by the transport equation
\begin{equation}
\partial_t W_t = \mathcal{C}_{\mathrm{diss}}(W_t) + \mathcal{C}_{\mathrm{cons}}(W_t)\,.
\end{equation}
The conservative part of the collision operator is defined through an effective hamiltonian,
\begin{equation}
\mathcal{C}_{\mathrm{cons}}(W_t)(k) = -i[H_\mathrm{eff}[W_t](k), W_t(k)]\,,
\end{equation}
where
\begin{equation*}
H_\mathrm{eff}[W](k)=\rm {pv}\int_{(\mathbb{T}^d)^3}dk_1dk_2dk_3\delta(k_0+k_1-k_2-k_3)\frac{1}{\omega}\times \big((1-W_2)J[W_1(1-W_3)]+W_2J[(1-W_1)W_3]\big)
\end{equation*}
with $\omega$ a collision energy, and pv standing for principal value around $\omega=0$. $\mathcal{C}_{\mathrm{cons}}$ is like a Vlasov term and does not generate any entropy. The dissipative part of the collision operator coincides with the one in Section 2a for the case $B\equiv 1$, and is defined by
\begin{eqnarray*}
\mathcal{C}_{\mathrm{diss}}(W)(k_0) =
\pi\int_{(\mathbb{T}^d)^3}dk_1dk_2dk_3\delta(k_0+k_1-k_2-k_3)\delta(\omega)\times
\big((1-W_0)W_2J[(1-W_1)W_3]\\
+J[W_3(1-W_1)]W_2(1-W_0)-W_0(1-W_2)J[W_1(1-W_3)]-J[(1-W_3)W_1](1-W_2)W_0\big).
\end{eqnarray*}
If we set $H_\mathrm{eff}[W] = 0$  by hand, the existence and uniqueness of solutions to (2.8) can be proved
following the blueprint [D]. The non-commutativity is controlled by the matrix inequality
\begin{equation}
 A J[BC] + C J[B A] \ge 0\,
\end{equation}
for arbitrary non-negative matrices $A,B,C$, see the recent preprint [LMS]. $H_\mathrm{eff}[W]$
turns out to be unbounded, since no a priori smoothness of $W$ is available and the principal value can become large or even ill-defined. In [LMS] it is explained how one can circumvent this difficulty and thereby arrive at a global existence result.
The solution preserves the Fermi property and conserves energy, $\int \mathrm{Tr}(W(k)) \omega(k) dk$,
and  ``spin'' $\int W(k)dk$. The H-theorem holds. Under suitable additional assumptions the stationary solutions are all
of Fermi-Dirac type and diagonal in the basis defined by the initial $\int W(k)dk$. For more detailed discussions and numerical solutions we refer to [FMS].\\
\\

\section{A two-component boson problem}
{\bf3 a) Low temperature bosons and a condensate. }\\
\\
\setcounter{equation}{0}
The phenomenon of Bose-Einstein condensation occurs when a large number of particles of a Bose gas enter the same lowest accessible quantum state. Predicted by Bose [B] and Einstein [Ei] in 1924, it was first unambiguously produced in 1995 by E. Cornell and C. Wieman.
We shall now discuss a Bose condensate below the transition temperature $T_c$ for condensation, and in interaction with a non-condensates component. The setting is a two-component model well established in physics (see the monograph [GNZ] and its references) of pair-collision interactions involving a gas of thermally excited (quasi-)particles and a condensate. The two-component model consists of a kinetic equation for the distribution function of the gas, and a Gross-Pitaevskii equation (cf [PS]) for the condensate.
In the superfluid frame a rather general form of the kinetic equation is (cf [PBMR], [ZNG])
\begin{eqnarray}
\partial _tf+ \nabla_p(E_p)+p_c)\cdot\nabla_x f-\nabla_x(E_p)\cdot\nabla_pf = Q_{NUU}(f)+C_{12}(f,n_c).
\end{eqnarray}
Here $f$ is the quasi-particle density, $n_c$ (resp. $p_c$) is the mass density (resp. the momentum) of the condensate, and $E_p$ denotes the (Bogoliubov) quasi-particle excitation energy,
\begin{eqnarray}
E_p= \sqrt{\frac{\lvert p\rvert ^4}{4m^2}+\frac{gn_c}{m}\lvert p\rvert ^2},
\end{eqnarray}
where $g= \frac{4\pi a\hbar ^2}{m}$, $a$ is the scattering length of the interaction potential, and
$m$ the atomic mass.
The Nordheim-Uehling-Uhlenbeck term $Q_{NUU}$ for collisions between (quasi-)particles is given by (1.1).
The collision term $C_{12}$ for collisions between (quasi-)particles and condensate is
\begin{eqnarray}
C_{12}(f,n_c)(p)=
\frac{g^2 n_c}{\hbar}\int_{\R ^3\times \R ^3\times \R ^3} |A|^2\delta(p_0+p_1-p_2-p_3)\delta(E_1-E_2-E_3)[\delta(p-p_1)\\
-\delta(p-p_2)-\delta(p-p_3)]
((1+f_1)f_2f_3-f_1(1+f_2)(1+f_3))dp_1dp_2dp_3,\nonumber
\end{eqnarray}
where
\begin{eqnarray*}
f_j=f(p_j),\quad E_j=E(p_j),\quad1\leq j\leq3.
\end{eqnarray*}
We notice that the domain of integration is only 2d and not 5d as for classical Boltzmann. The transition probability kernel $|A|^2$ can be explicitly computed
by the Bogoliubov approximation scheme.
\begin{eqnarray*}
A:= (u_3-v_3)(u_1u_2+v_1v_2)+(u_2-v_2)(u_1u_3+v_1v_3)-(u_1-v_1)(u_2v_3+v_2u_3).
\end{eqnarray*}
Here the Bose coherence factors $u$ and $v$ are
\begin{eqnarray*}
u_p^2= \frac{\tilde{\ep }_p+E_p}{2E_p},\quad u_p^2-v_p^2= 1.
\end{eqnarray*}
\\
The collision operator $C_{12}(f,n_c)$ can be formally obtained (cf [ST], [EMV], [No]) from the
Nordheim-Boltzmann collision operator
\begin{eqnarray}
Q_{NUU}(f)(p)= \int _{\R ^3\times \R ^3\times \R ^3}B\delta (p+p_*-( p'+p'_*))\delta (E(p)+E(p_*)-( E(p')+E(p'_*)))\nonumber\\
\Big( f'f'_*(1+f)(1+f_*)-ff_*(1+f')(1+f'_*)\Big) dp_*dp'dp'_*.
\end{eqnarray}
Namely, assume that a condensate appears below the Bose-Einstein condensation temperature $T_ {c}$. That splits the quantum gas distribution function into a condensate part $n_c\delta _{p= 0}$ and an $L^1$-density part $f(t,x,p)$, and gives
\begin{eqnarray*}
Q_{NUU}(f+n_c\delta _{p= 0})= Q_{NUU}(f)+C_{12}(f,n_c)+b n_c^2 +c n_c^3 +d n_c\delta_{p=0},
\end{eqnarray*}
where a simple computation shows that $b= c= 0$.\\
In equilibrium, the right hand side of (3.1) vanishes. Multiplying the collision term by $\log\frac{ f}{1+ f}$ and integrating in $p$, it follows that in equilibrium
\begin{eqnarray}
\frac{f'}{1+f'}\frac{f'_*}{1+f'_*}
=\frac{f}{1+f}\frac{f_*}{1+f_*}, \quad {\rm when}\quad p+p_*=p'+p'_*, \quad \lvert p\rvert ^2+\lvert p_*\rvert ^2= \lvert p'\rvert ^2+\lvert p'_*\rvert ^2,\hspace*{0.2in}\\
\frac{f_1}{1+f_1}=\frac{f_2}{1+f_2}\frac{f_3}{1+f_3}, \quad \quad \quad \quad {\rm when}\quad p_1=p_2+p_3, \quad \lvert p_1\rvert ^2= \lvert p_2\rvert ^2+\lvert p_3\rvert ^2+n_c.\quad \quad \quad
\end{eqnarray}
Equations (3.5-6) imply that $\frac{f}{1+f}$ is a Maxwellian, hence $f$ a Planckian, which is of the type
\begin{eqnarray*}
\frac{1}{e^{\alpha(\lvert p\rvert ^2+n_c)+\beta\cdot p}-1}, \quad\alpha>0,\quad\beta\in\mathbb{R}^3,\quad p\in \R ^3.
\end{eqnarray*}
At the low temperatures we have in mind here, the number of excited (quasi-)particles are considered to be small, when they are sufficiently excited for pair collisions to be important. Also the time-scale to reach equilibrium for such collisions, is considered to be short compared to the one for $C_{12}$, and so the $Q_{NUU}$ collision term is neglected relative to the collision operator $C_{12}$. \\
\hspace{1cm}\\
The usual Gross-Pitaevskii (GP) equation for the wave function $\psi$ (the order parameter) associated with a Bose condensate is
\begin{eqnarray*}
ih\frac{\partial \psi}{\partial t}=-\frac{h^2}{2m}\Delta_x\psi+(U_{ext}+g|\psi|^2)\psi,
\end{eqnarray*}
i.e. a Schr\"odinger equation with $U_{ext}$ an external potential, and complemented by a non-linear term accounting for two-body interactions.
For simplicity, we do not include the strongly space-inhomogeneous trapping potential, otherwise an essential ingredient in experiments on laser-trapped Bose gases.
Modulo a numerical factor, $g$ is the $s$-scattering length of the two-body interaction potential. In the present context the Gross-Pitaevskii equation is further generalized by letting the condensate move in a self-consistent (Hartree-Fock) mean field $2g\tilde{n}=2g\int f(p)dp$ produced by the thermally excited atoms, together with a dissipative coupling term associated with the collisions.
The generalized Gross-Pitaevskii equation derived in e.g. [KD1], [KD2], [PBMR] and [[GNZ], is of the type
\begin{eqnarray}
i\hbar \partial _t\psi= &-\frac{\hbar^2}{2m}\Delta_x\psi+\Big( g|\psi|^2+U_{ext}+2g\int_{\R ^3} fdp+i\frac{g^2}{2\hbar}\int_{\R ^3\times \R ^3\times \R ^3} \delta(p_1-p_2-p_3)\delta(E_1-E_2-E_3)\nonumber\\
&((1+f_1)f_2f_3-f_1(1+f_2)(1+f_3))dp_1dp_2dp_3\Big) \psi.
\end{eqnarray}
\\
\setcounter{theorem}{0}
{\bf3 b) A space-homogeneous, isotropic case.}\\
\\
Consider the Cauchy problem for the two component model (3.1), (3.7) in a space-homogeneous isotropic situation and in the superfluid rest
frame (condensate velocity $v_s=\bigtriangledown_x \theta =0$), i.e. the equations
\begin{eqnarray}
\frac{\partial f}{\partial t} = C_{12}(f,n_c),\\
\frac{d n_c}{d t}=-\int C_{12}(f,n_c)dp,
\end{eqnarray}
with initial values
\begin{eqnarray}
f(p,0)=f_i(\mid p\mid), \quad n_c(0)=n_{ci}.
\end{eqnarray}
Here $f(p,t)$ is the density of the quasi-particles, $n_c(t)$ the mass of the condensate, and the collision operator $C_{12}$ is given by (3.3) in the local rest frame.\\
\hspace{1cm}\\
There are two physically important regimes (cf [E]). One is the very low temperature situation with all $|p_i|<<p_0$ ($T\leq 0.4T_c$ in the set-up of [E]), i.e. where physically all quasi-particle momenta are much smaller than the characteristic momentum $p_0= \sqrt{2mgn_c}$ for the crossover between the linear and quadratic parts of the Bogoliubov excitation energy of the quasi-particles;
\begin{eqnarray}
E(p):= \sqrt{\frac{p^4}{4m^2}+\frac{gn_c}{m}p^2}\approx c|p|(1+\frac{p^2}{8gmn_c})=c|p|(1+\frac{p^2}{4p_0^2}).
\end{eqnarray}
Here $c:= \sqrt{\frac{gn_c}{m}}$ the speed of Bogoliubov sound.  Setting $m=\frac{1}{\sqrt 2}$ gives $p_0=c$. In applications with $|p| << p_0$, the right hand side of (3.11) is usually taken as the value of $E(p)$.\\
The Bose coherence factors can then be taken as
\begin{eqnarray*}
u_p= \sqrt{\frac{gn_c}{2E(p)}}+\frac{1}{2}\sqrt{\frac{E(p)}{2gn_c}}, \quad v_p= \sqrt{\frac{gn_c}{2E(p)}}-\frac{1}{2}\sqrt{\frac{E(p)}{2gn_c}},\quad u_p^2-v_p^2=1,
\end{eqnarray*}
which gives
\begin{eqnarray*}
A= \frac{1}{2^\frac{5}{2}}\frac{\sqrt{E(p_*)E(p')E(p'_*)}}{(gn_c)^\frac{3}{2}}
+\sqrt{\frac{gn_c}{2}}(\sqrt{\frac{E(p'_*)}{E(p_*)E(p')}}+\sqrt{\frac{E(p')}{E(p_*)E(p'_*)}}
-\sqrt{\frac{E(p_*)}{E(p')E(p'_*)}}).
\end{eqnarray*}
And so recalling that $ E(p_*)= E(p')+E(p'_*)$, we obtain
\begin{eqnarray*}
A=  \frac{1}{2^\frac{5}{2}}\frac{\sqrt{E(p_*)E(p')E(p'_*)}}{(gn_c)^\frac{3}{2}}.
\end{eqnarray*}
With this $A$, the collision operator becomes
\begin{eqnarray}
C_{12}(f,n_c)(p)= \int \chi \frac{{E(p_1)E(p_2)E(p_3)}}{g^3n_c^2}\delta(p_1-p_2-p_3)\delta(E_1-E_2-E_3)[\delta(p-p_1)\nonumber \\
-\delta(p-p_2)-\delta(p-p_3)]
((1+f_1)f_2f_3-f_1(1+f_2)(1+f_3 ))dp_1dp_2dp_3,
\end{eqnarray}
where $ \chi $ denotes the truncation for ${|p_i|\leq \lambda}$, $1\leq i\leq 3$ for a given $\lambda>0$.\\
\hspace*{1.in}\\
The opposite limit where all momenta $|p_i|>>p_0$ has the dominant excitation of (Hartree-Fock) single particle type (in [E] for moderately low temperatures around $0.7T_{c}$). Expanding the square root definition of $E$ in (3.11), we may approximate $E_p$ by $\frac{p^2}{2m}+gn_c$ leading to a collision operator of the type
\begin{eqnarray}
C_{12}(f,n_c)= k n_c\int_{\R ^3\times \R ^3\times \R ^3}  \chi \delta(p_1-p_2-p_3)\delta(E_1-E_2-E_3)[\delta(p-p_1)\nonumber \\
-\delta(p-p_2)-\delta(p-p_3)]
((1+f_1)f_2f_3-f_1(1+f_2)(1+f_3 ))dp_1dp_2dp_3
\end{eqnarray}
for the 'partial local equilibrium regime'. Here $\chi $ is the characteristic function of the set of $(p, p_1, p_2, p_3)$ with $|p|$, $|p_1|$, $|p_2|$, $|p_3|\geq \alpha $ for a given positive constant $\alpha $.\\
\hspace*{1.in}\\
In the general case, the collision operator is
\begin{eqnarray}
C_{12}(f,n_c)(p)= kn_c\int |A|^2\delta(p_1-p_2-p_3)\delta(E_1-E_2-E_3)[\delta(p-p_1)\nonumber\\
-\delta(p-p_2)-\delta(p-p_3)]
((1+f_1)f_2f_3-f_1(1+f_2)(1+f_3))dp_1dp_2dp_3,
\end{eqnarray}
with the excitation energy $E$ defined by
\begin{eqnarray*}
E(p)= |p|\sqrt{\frac{p^2}{4m^2}+\frac{gn_c}{m}}.
\end{eqnarray*}
The kernel $|A|^2$ is bounded by a multiple of
\begin{eqnarray*}
|\bar{A}|^2:= \Big( \frac{|p_1|}{\sqrt{n_c}}\wedge 1\Big) \Big( \frac{|p_2|}{\sqrt{n_c}}\wedge 1\Big) \Big( \frac{|p_3|}{\sqrt{n_c}}\wedge 1\Big) ,
\end{eqnarray*}
in the physically interesting cases where asymptotically all $|p_i|<<p_0$, all $|p_i|>>p_0$, or one $|p_i|<<p_0$ and the others $>>p_0$. These three cases are relevant for very low respectively moderately low temperatures compared to $T_{c}$, and (the third case) for the collision of low energy phonons with high energy excitations (atoms). The asymptotic situation of two $|p_i|<<p_0$ and one $p_i>>p_0$  (with unbounded $A$) is excluded by the energy condition. \\
Using $|\bar{A}|^2$ as the kernel in the collision operator, the following result holds.
\begin{theorem}
{\rm [AN1]} Let $n_{ci}>0$ and $f_i(p)= f_i(|p|) \in L^1$ be given with $f_i$ nonnegative and $f_i(p)|p|^{2+\gamma} \in L^1$ for some $\gamma>0$. For the collision operator (3.14) with the transition probability kernel $|\bar{A}|^2$, there exists a nonnegative solution $(f,n_{c})\in  C^1([0,\infty);L^1_+)\times C^1([0,\infty))$ to the initial value problem (3.1), (3.7) in the space-homogeneous, isotropic case. The condensate density $n_c$ is locally bounded away from zero for $t>0$. The excitation density $f$ has energy locally bounded in time. Total mass $M_0=n_{ci}+\int f_i(p)dp$ is conserved, and the moment $\int |p|^{2+\gamma}fdp$ is locally bounded in time. In the moderately low temperature case a total energy quantity is conserved.
\end{theorem}
The proof is via approximations controlled by a priori estimates and fixed point techniques. An important still open problem is the question of time-asymptotics. The addition of a NB collision term would (using methods we are aware of) considerably weaken the type of solutions that can be obtained.\\

\hspace*{1.in}\\
The paper [S2] considers the spatially homogeneous and isotropic kinetic regime of weakly interacting bosons with s-wave scattering. It has a focus on post-nucleation self-similar solutions. Another paper, [EPV], studies linearized space homogeneous kinetic problems in settings related to the ones discussed here, and with a focus on large time behaviour.

\hspace*{1.in}\\
{\bf3 c) A space-dependent, close to equilibrium case.}\\
\\
We shall discuss the stability of an equilibrium of the system in a periodic slab, under small deviations that respect the conservation laws. In equilibrium the density of the excitations is
 $\frac{1}{(e^{\alpha(p^2+n_c)+\beta\cdot p}-1}, \quad\alpha>0,\quad\beta\in\mathbb{R}^3 $.
We restrict to $|p|>>\sqrt{2mgn_c}$, and more particularly consider
the moderately low temperature range T close to $0.7T_{c}$, where the approximation $\frac{ p^2}{  2m} +gn_c$ of $E(p)$ is  commonly used.
Set $\alpha=1$,  fix the equilibrium limit for $\psi$ of (3.3) as $n_c=n_0>0$ and take $n_c=n_0 $ in the equilibrium Planckian, set $|\beta|=2\sqrt{2mgn_0}$, and write the Planckian as $\frac{1}{e^{(p-p_0)^2}-1}$ with $p_0=-\frac{\beta}{2}$. Changing variables $p\rightarrow p-p_0$ gives $P:=\frac{1}{e^{ p^2}-1}$ as Planckian equilibrium for $f$.\\
Taking into account that the system is close to equilibrium, introduce a mean free path $\epsilon$, so that $C_{12}$ becomes $\frac{1}{\epsilon }C_{12}$. The factor $g$ is proportional to the scattering length, which is smaller than the mean free path$\epsilon$. Take $\lambda=(\frac{g}{\epsilon})^2<1$. The functions $(f(t,x,p),\psi(t,x))$ are considered in the slab $\Omega = [ 0,2\pi ] $ in the $x$-direction with periodic boundary conditions. They are small deviations of the equilibrium,  $(f(t,x,p),\psi(t,x))=(P(1+\lambda R),\sqrt{n_0}+\lambda\Phi)$.\\
The external potential $U_{ext}$ is assumed to be a constant that will be fixed later.
Taking $m= \frac{1}{2}$, $\hbar=1$ for the sake of simplicity, the system of equations to be satisfied by $(f,\psi )$ becomes
\[ \begin{aligned}
&\partial _tf+p_x\partial _xf= {g\sqrt{\lambda}\lvert \psi \rvert ^2}\int _{\R ^3\times \R ^3\times \R ^3}\delta_3\delta(p_1-p_2-p_3)\delta(E_1-E_2-E_3)(f_2f_3-f_1(1+f_2+f_3))
dp_1dp_2dp_3,\\
&\partial _t\psi -i\partial ^2_x\psi
= \Big( \frac{\sqrt{\lambda}}{2}\int_{\R ^3\times \R ^3\times \R ^3} \delta(p_1-p_2-p_3)\delta(E_1-E_2-E_3)\\
&\hspace*{1.7in}(f_2f_3-f_1(1+f_2+f_3))dp_1dp_2dp_3
-i(\lvert \psi \rvert ^2+\frac{U_{ext}}{g}+2\int_{\R ^3} fdp)\Big)g \psi ,
\end{aligned}\]
\hspace{1cm}\\
where $\delta_3=\delta (p- p_1)-\delta (p- p_2)-\delta (p- p_3)$.\\
The function $f$ is taken cylindrically symmetric in $p= (p_x,p_r)\in \R \times \R ^2$
. This changes the linear moment conservation Dirac measure in the collision term to $\delta(p_{1x}-p_{2x}-p_{3x})$. In our temperature range, as explained in [E], [KD1-2], [ZNG] and more in details in [ITG], the factor $|A|^2$ in the collision term is usually  taken as one, so that, in the sequel, we set $|A|^2=1$. Since the collective excitations play no role in this temperature range, the domain of integration is here taken as the set of $p\in \R^3$ such that $\lvert p\rvert^2 >2\Lambda^2 $ with $\Lambda >2\sqrt{gn_0}$. Denote by $\tilde{\chi}$ the characteristic function of the set
\begin{eqnarray*}
\{ (p,p_1,p_2,p_3)\in \R ^3\times \R ^3\times \R ^3\times\R^3;|p|^2, |p_1|^2,|p_2|^2, |p_3|^2>2\Lambda^2\}.
\end{eqnarray*}
The restriction $|p|^2>2\Lambda^2$ will be implicitly assumed below, and $\int dp$ will stand for $\int _{|p|^2>2\Lambda^2} dp$.
Set
$\delta _0= \delta (p_{1x}= p_{2x}+p_{3x}, \quad \lvert p_1\rvert ^2= \lvert p_2\rvert ^2+\lvert p_3\rvert ^2+gn_0)$,
and consider the system
\begin{equation}\label{eq-f}
\partial _tf+p_x\partial _xf= {g\sqrt{\lambda} n_c}\int _{\R ^3\times \R ^3\times \R ^3}\tilde{\chi}\delta _0\delta _3(f_2f_3-f_1(1+f_2+f_3))dp_1dp_2dp_3,
\end{equation}
\begin{equation}\label{init-f}
f(0,x,p)=f_i(x,p),\nonumber
\end{equation}
and
\begin{equation}\label{eq-psi}
\partial _t\psi -{i}\partial ^2_x\psi \\
= \Big( \frac{\sqrt{\lambda}}{2}\int_{\R ^3\times \R ^3\times \R ^3} \tilde{\chi}\delta _0(f_2f_3-f_1(1+f_2+f_3))dp_1dp_2dp_3-i(n_c+\frac{U_{ext}}{g}+2\int fdp)\Big) g\psi ,
\end{equation}
\begin{equation*}
\psi (0,x)=\psi_i(x).
\end{equation*}
Here, the function $n_c$ is defined by $n_c=n_c(t,x):= |\psi|^2(t,x)$.
The total initial mass is
\begin{eqnarray*}
2\pi \mathcal{M}_0:= \int _\Omega |\psi_i(x)|^2dx+ \int  _{\Omega \times \R ^3}f_i(x,p)dxdp,
\end{eqnarray*}
which is formally conserved by the equations (3.15-16).\\
The initial data $f_i$ and $\psi _i$ are taken as
\begin{eqnarray*}
f_i:= P(1+\lambda R_i),\quad \psi_i:  = \sqrt{n_0}+\lambda \Phi _i,
\end{eqnarray*}
for some functions $R_i(x,p)$ and $\Phi _i(x)$ with
\begin{equation*}\label{cond2-Ri}
\int (|\psi_i|^2-n_0+\lambda\int_{\mathbb{R}^3}{P}R_idp)dx=0.
\end{equation*}
This is consistent with the asymptotic behavior proven in the paper, i.e. $(f-P,\lvert \psi \rvert ^2-n_0)$ tending to zero when time tends to infinity.
It implies that (up to the multiplicative constant $\frac{1}{2\pi }$) the initial (and conserved) total mass equals the mass of $(P,n_0)$, i.e.
\begin{equation}\label{cons-mass}
\mathcal{M}_0= \int P(p)dp+n_0.
\end{equation}
The separate masses of condensate and excitation may, however, not be conserved.
The constant $U_{ext}$ will be taken as $g(n_0-2\mathcal{M}_0)$. \\
\hspace{1cm}\\
The main results concern the well-posedness and long time behaviour of the problem (3.15-16).\\
Let $\parallel.\parallel_2$ denote the norm in $L^2([0,2\pi])$, and
set $\parallel\psi\parallel_{H^1}:=\parallel\psi\parallel_2+
\parallel\partial_x\psi\parallel_2$, \\
let  $\parallel.\parallel_{2,2}$ denote the norm in $L^2_{\frac{P}{1+P}}([0,2\pi]\times\mathbb{R}^3)$, i.e. $\parallel h\parallel _{2,2}:= (\int h^2(x,p)\frac{P}{1+P}dpdx)^\frac{1}{2}$, \\
and let $L^2_{\frac{1}{P(1+P)}}$ denotes the $L^2$-space of functions $h$ with norm $(\int \frac{h^2(x,p)}{P(1+P)}dpdx)^\frac{1}{2}$.\\
The solutions of (3.15) will be strong solutions, i.e. such that the collision operator $C_{12}(f,n_c)$ belongs to $C_b\big( \R ^+; L^2_{\nu ^{-\frac{1}{2}}\sqrt{\frac{P}{1+P}}}(\R ^3;H^1(0,2\pi ))\big) $, $\nu $ being the collision frequency. The solutions of (3.16) are $H^1$-solutions in the following sense. A function $\psi\in \mathcal{C}_b(\mathbb{R}^+;H_{{\rm per}}^1(0,2\pi)$ is an $H^1$-solution to (3.16), if for all $\phi\in \mathcal{C}(\mathbb{R}^+;H_{{\rm per}}^1(0,2\pi))$ and all $t>0$,
\[ \begin{aligned}
&\int \psi(t,x)\bar{\phi}(t,x)dx-\int \psi_i(x)\bar{\phi}(0,x)dx+{i}\int_0^tds \int dx\partial_x\psi(s,x)\partial_x\bar{\phi}(s,x) \\
&=\int_0^t ds\int dx\Big( \frac{\sqrt{\lambda}}{2}\int_{\R ^3\times \R ^3\times \R ^3} \tilde{\chi}\delta _0(f_2f_3-f_1(1+f_2+f_3))dp_1dp_2dp_3-i(n_c+n_0-2\mathcal{M}_0+2\int fdp+U_{ext})\Big) g\psi\bar{\phi}.
\end{aligned}\] \\
\begin{theorem}
{\rm[AN3]} There are $\lambda_1,c_\zeta\in]0,1[$, $\eta_1>0$, and $K>0$, such that for $\lambda<\lambda_1$, and \\
 $(R_i,\Phi_i)\in L_{(1+|p|)^3\frac{P}{1+P}}^2(\mathbb{R}^3;H^1_{{\rm per}}(0,2\pi))\times H_{{\rm per}}^1(0,2\pi)$ with
\begin{eqnarray}\label{cond1-Ri}
\int R_i(x,p)p_x{P}dxdp= \int R_i(x,p)(|p|^2+gn_0){P}dxdp= 0,
\end{eqnarray}
\begin{equation}\label{cond2-Ri}
\int (|\psi_i|^2-n_0+\lambda\int_{\mathbb{R}^3}{P}R_idp)dx=0,
\end{equation}
and
\begin{eqnarray}\label{cond3-Ri}
\parallel \Phi _i\parallel _{H^1}\leq \eta _1,\quad \parallel R_i\parallel _{2,2}+\parallel \partial _xR_i\parallel _{2,2}\leq \eta _1,
\end{eqnarray}
there is a unique solution
\begin{eqnarray*}
(f,\psi)=(P(1+\lambda R),\sqrt{n_0}+\lambda\Phi)\in  \mathcal{C}_b(\mathbb{R}^+;L^2_{\frac{1}{P(1+P)}}(\R^3;H_{{\rm per}}^1(0,2\pi)))\times \mathcal{C}_b(\mathbb{R}^+;H_{{\rm per}}^1(0,2\pi))
\end{eqnarray*}
to (3.15-16) with $f>0$. For all $t\in \mathbb{R}^+$, the solution satisfies,
\begin{eqnarray}\label{bounds-Phi-s}
f\in L^2_{\frac{(1+|p|)^3}{P(1+P)}}([0,t]\times\R^3;H_{{\rm per}}^1(0,2\pi))),\nonumber\\
\parallel R(t,\cdot ,\cdot )\parallel _{2,2}+\parallel \partial _xR(t,\cdot ,\cdot )\parallel _{2,2}\leq K\eta _1e^{-\zeta t},\\
\int (\lvert \partial _x\psi\rvert ^2+\frac{g}{2}(\lvert \psi\rvert ^2-n_0)^2)(t,x)dx
\leq K\lambda\nonumber,
\end{eqnarray}
where $\zeta=c_\zeta \sqrt{\lambda}$. Moreover, $n_c(t)=\int|\psi(t ,x)|^2dx$ converges exponentially of order $\zeta$ to $n_0$ when $t\rightarrow +\infty$,
\begin{equation*}
\lim_{t\rightarrow +\infty}\int (\lvert \partial _x\psi\rvert ^2+\frac{\epsilon ^2}{2}(\lvert \psi\rvert ^2-n_0)^2)(t,x)dx
\end{equation*}
 exists, and the convergence to its limit is exponential of order $\zeta$.
\end{theorem}
\[\]
Whereas non-linear systems of the type (3.15-16) and its generalizations have been much studied in mathematical physics below $T_c$, there are so far only few papers with their focus mainly on the non-linear mathematical questions. Starting from a similar Gross-Pitaevskii and kinetic frame, two-fluid models are derived in [A]. A Milne problem related to the present set-up is studied in [AN2]. The paper [EPV] considers a related setting,  and has its focus on linearized space homogeneous problems. Validation aspects in the space-homogeneous case are discussed in [S2].
There has been a considerable interest recently (see ee.g. [EV1-2], [Lu4] and references therein) in the bosonic Nordheim-Uehling-Uhlenbeck equation as a model above and around $T_c$ for blow-ups and for condensation in space-homogeneous boson gases.
\\
A classical approach to study kinetic equations in a perturbative setting, is to use a spectral inequality (resp. Fourier techniques and the $\parallel \cdot \parallel _{T,2,2}$ norm) for controlling the non-hydrodynamic (resp. hydrodynamic) parts of the solutions. An additional problem here is the coupling with the generalized Gross-Pitaevskii equation. The general approach, together with a Fourier based analysis of the generalized Gross-Pitaevskii equation, provide local in time solutions to the present coupled system.
Since the condensate and the normal gas are coupled by the collision interaction, the exponential decrease of
the deviation of the kinetic distribution function from the equilibrium Planckian $P$, helps to control the long-term evolution of the condensate. This is an important ingredient in the passage from local to global solutions,
which also leads to exponential decreases of the deviation of the condensate mass from its equilibrium state ${n_0}$, and of the energy in (3.21) from its limit value, even though there are bounds for, but not conservation of the total energy in the model.\\
Within this frame, the kinetic equation (3.15) differs from earlier classical ones. The collision operator in space-homogeneous bosonic Nordheim-Uehling-Uhlenbeck papers has so far been taken isotropic, but is here, due to the space-dependent slab-context, cylindric. Mass density does not belong to the kernel of the present linearized collision operator, and the scaling at infinity in its collision frequency is stronger than in the classical case.\\
\\
The one-dimensional spatial frame induces simplifications of the functional analysis, mainly due to the control of the condensate.
The $\T ^d$ spatial frame, for $d\geq 2$, is still open. \\
The conservation properties of the model (3.15-16), as well as some properties of the collision operator $\frac{C_{12}}{n_c}$ and its linearized operator $L$ around the Planckian $P$, are used to obtain a priori estimates for some linear equations related to (3.15) and (3.16). These are used in the proof of the main theorem. The proof starts with a contractive iteration scheme to obtain local solutions. A key point in the global analysis is the consequences of the exponential convergence to equilibrium for $f$. The analysis of $\psi$ differs from the classical Gross-Pitaevskii case in using the exponential convergence to equilibrium of $f$ to control the behaviour of the kinetic energy $\frac{1}{2\epsilon^2}\int |\partial_x\psi|^2dx$  and the internal energy $\frac{1}{2}\int |\psi|^4dx$ of $\psi$, in the proof of global in time solutions.\\
\\

\section{A kinetic anyon equation.}
\setcounter{theorem}{0}
\setcounter{equation}{0}
{\bf 4 a) Anyons and the Boltzmann equation.}\\
\\
To recall the definition of anyon, consider the wave function $\psi(R,\theta,r,\varphi)$ for two identical particles with center of mass coordinates $(R,\theta)$ and relative coordinates $(r,\varphi)$. Exchanging them,
$\varphi\rightarrow \varphi+\pi$, give a phase factor $e^{2\pi i}$  for bosons and $e^{\pi i}$ for fermions. In three or more dimensions those are all possibilities. Leinaas and Myrheim proved in 1977 [LM], that in one and two dimensions any phase factor is
possible in the particle exchange. The new (quasi-)particles became a hot topic after the the first experimental confirmations in the early 1980-ies, and Frank Wilczek in analogy with the terms bos(e)-ons and fermi-ons coined the name any-ons  for the new (quasi-)particles with any phase. Anyon quasi(-)particles with e.g. fractional electric charge, have since been observed in various types of experiments.\\
By moving to a definition in terms of a generalized Pauli exclusion principle, Haldane [H] generalized this  to a fractional exclusion statistics valid for any dimension, and coinciding with the anyon definition in the one and two dimensional cases. Haldane statistics has also been realized for neutral fermionic atoms at ultra-low temperatures in three dimensions. Wu later derived [W] occupation-number distributions for ideal gases under Haldane statistics by counting states under the new fractional exlusion principle.
From the number of quantum states  of $N$ identical particles occupying $G$ states being
\begin{eqnarray*}
\frac{(G+N-1)!}{ N!(G-1)!} \hspace{.5cm} {\rm and}\hspace{.5cm} \frac{G!}{N!(G-N)!}
\end{eqnarray*}
in the boson resp. fermion cases, he derived the interpolated number of quantum states for the fractional exclusions to be
\begin{eqnarray}
\frac{(G+(N-1)(1-\alpha))!}{N!(G-\alpha N-(1-\alpha))!}\hspace{.1cm},\hspace{1cm} 0<\alpha<1.
\end{eqnarray}
 This led him to to an equilibrium statistical distribution for ideal fractional exclusion gases
\begin{eqnarray}
\frac{1}{w(e^{(\epsilon-\mu)/T})+\alpha}\hspace{.1cm},
\end{eqnarray}
where $\epsilon$ is particle energy, $\mu$ chemical potential, $T$ temperature, and the function $w(\zeta)$ satisfies
\begin{eqnarray*}
w(\zeta)^\alpha(1+w(\zeta))^{1-\alpha}=\zeta\equiv e^{(\epsilon-\mu)/T}.
\end{eqnarray*}
In particular $w(\zeta)=\zeta-1$ for $\alpha=0$ (bosons) and $w(\zeta)=\zeta)$ for $\alpha=1$ (fermions).
\hspace{1cm}\\
An important question for gases with fractional exclusion statistics, is how to calculate their transport properties, in particular how the Boltzmann equation
 \begin{equation*}
\partial _tf+v\bigtriangledown_xf= Q(f)
\end{equation*}
 gets modified.
 An answer was given by Bhaduri and coworkers [BBM] by generalizing to anyons the filling factors from the fermion and boson cases. With a filling factor $F(f)$  in the collision operator $Q$, the entropy production term becomes
\begin{eqnarray*}
\int Q(f)\log\frac{f}{F(f)}\hspace{.1cm},
\end{eqnarray*}
which for equilibrium implies
\begin{eqnarray*}
\frac{f'}{F(f')}\frac{f'_*}{F(f'_*)}=\frac{f}{F(f)}\frac{f_*}{F(f_*)}\hspace{.1cm}.
\end{eqnarray*}
Elastic pair collisions preserve mass, linear momentum, and energy in a Boltzmann type collision operator.
Using the conservation laws, we conclude as usual, that in equilibrium
\begin{eqnarray*}
\frac{f}{F(f)}=M
\end{eqnarray*}
is a Maxwellian.  Inserting Wu's equilibrium, this gives
\begin{eqnarray*}
f=\frac{1}{w(e^{(\epsilon-\mu)/T})+\alpha}\hspace{.1cm},\quad \quad F(f)=f e^{(\epsilon-\mu)/T}=\frac{e^{(\epsilon-\mu)/T}}{w(e^{(\epsilon-\mu)/T})+\alpha}\hspace{.1cm}.
\end{eqnarray*}
That is consistent with taking an interpolation between the fermion and boson factors as general filling factor, $F(f)= (1-\alpha f)^{\alpha}(1+(1-\alpha)f)^{1-\alpha}$, $0<\alpha<1$.
It gives as collision operator $Q$ under Haldane statistics [BBM] ,
\begin{eqnarray}
Q(f)= \int_{I\! \!R^d \times S^{d-1}}B(v-v_*,\omega)
\times [f'f'_*F(f)F(f_*)-ff_*F(f')F(f'_*)] dv_*d\omega.
\end{eqnarray}
Here $d\omega$ corresponds to the Lebesgue probability measure on the $(d-1)$-sphere. The collision kernel $B(z,\omega)$ in the variables $(z,\omega)\in I\! \!R^d \times \mathbb{S}^{d-1}$ is positive, locally integrable, and only depends on $|z|$ and $|(z,\omega)|$. See the discussion in [BBM] on the further choice of the kernel $B$ . The anyon Boltzmann equation for $0<\alpha<1$ retains important properties from the Fermi-Dirac case, but it has so far not been validated from basic quantum theory. \\
$F$ is concave with maximum value at $f=0$ for $\alpha \geq \frac{1}{2}$, and maximum value
at $f= \frac{1-2\alpha}{\alpha(1-\alpha)}$ for $\alpha <\frac{1}{2}$. In the filling factor $F(f)= (1-\alpha f)^{\alpha}(1+(1-\alpha)f)^{1-\alpha}$, $0<\alpha<1$, the factor $(1-\alpha f)^{\alpha}$ requires the value of $f$ to be between $0$ and $\frac{1}{\alpha}$.
This is formally preserved by the equation, since the gain term vanishes for $f=\frac{1}{\alpha}$, making the $Q$-term and the derivative left hand side of the Boltzmann equation negative there. And the derivative equals the positive gain term for $f=0$, where the loss term vanishes.\\
The collision  operator vanishes identically for the equilibrium distribution functions obtained by Wu, and for certain Heaviside functions.
For the limiting Nordheim-Boltzmann cases of boson ($\alpha=0$) and fermion statistics ($\alpha=1$), the quartic terms in the collision integral cancel, which is used when analysing them. 
But for $0<\alpha<1$ there are { no cancellations} in the collision term. Moreover, the Lipschitz continuity of the collision term in the Fermi-Dirac case, is replaced by a {weaker H\"older continuity} near $f=\frac{1}{\alpha}$. So when it comes to existence, weak $L^1$-methods seem to be excluded by the filling factors, and the available existence proofs use strong $L^1$ techniques.\\
\hspace{1cm}\\
\\
\setcounter{theorem}{0}
{\bf 4 b) The space-homogeneous anyon Cauchy problem.}\\
\\
The space-homogeneous initial value problem for the Boltzmann equation with Haldane statistics is
\begin{eqnarray}
\frac{ df}{dt}=Q(f),\quad f(0,v)=f_0(v).
\end{eqnarray}
Because of the filling factor $F$, the range for the initial value $f_0$ should belong to  $[0,\frac{1}{\alpha}]$, which is then formally preserved by the equation. A good control of $\int f(t,x,v)dv$, which in the space-homogeneous case is given by the mass conservation, can be used to also keep $f$ uniformly away from $\frac{1}{\alpha}$, and make $F(f)$ Lipschitz continuous. That was a basic idea behind the first existence result for the anyon Boltzmann equation.
\setcounter{theorem}{0}
\begin{proposition}
{\rm [A3]} Consider the space-homogeneous equation (4.4) with velocities in $I\! \!R^{d}$, $d\geq 2$ and for hard force kernels with
\begin{eqnarray}
0<B(z,\omega)\leq C|z|^{\beta} |\sin \theta \cos \theta |^{d-1},
\end{eqnarray}
where $0<\beta \leq 1$, $d>2$, and $0<\beta<1$, $d=2$. Let the initial value $ f_0$ have finite mass and energy. If $0<f_0\leq\frac{1}{\alpha}$ and ${\rm ess\hspace{.1cm}sup}(1+|v|^s)f_0 <\infty$ for $s=d-1+\beta$. Then the initial value problem for (4.4)
has a strong solution in the space of functions continuous from $t\geq 0$ into $L^1\cap L^{\infty}$, which conserves mass and energy, and for $t_0>0$ given, has ${\rm ess\hspace{.1cm}sup}_{v,t\leq t_0}|v|^{s'}f(t,v)$ bounded,where $s'=\min(s,\frac{2\beta(d+1)+2}{d})$.
\end{proposition}
\underline{\bf{Remarks.}} In this proposition, stronger limitations on $B$ would allow for weaker conditions on the initial value $f_0$.\\
The proof implies stability; (given a sequence of initial values  $(f_{0n})_{n\in\mathbb{N}}$ with
\begin{eqnarray*}
\sup_n {\rm esssup_v}\hspace{.2cm} f_{0n}(v) <\frac{1}{\alpha},
\end{eqnarray*}
and converging in $L^1$  to $f_0$, there is a subsequence of the solutions converging in $L^1$ to a solution with initial value $f_0$.)\\
What do the finite time blow-up solutions of [EV1-2] infer about the Haldane solutions, when $\alpha$ tends to zero?\\
\\
\underline{Idea of proof.}\\
This initial value problem is first considered for a family of approximations with bounded support for the kernel $B$, when $0<f_0\leq {\rm esssup} f_0<\frac{1}{\alpha}$. Starting from approximations with Lipschitz continuous filling factor, the corresponding solutions are shown to stay away uniformly from $\frac{1}{\alpha}$, the upper bound for the range. Uniform Lipschitz continuity follows for the approximating operators and leads to well-posedness for the limiting problem. Uniform $L^{\infty}$ moment bounds for the approximate solutions hold, using an approach from the classical Boltzmann case [A1]. Based on those preliminary results, the global existence  of the limit solutions follows by strong compactness arguments. The existence result for $f_0\leq \frac{1}{\alpha}$ finally follows by an initial layer analysis.\\
Mass and first moments are conserved and energy is bounded by its initial value. That bound on energy in turn implies energy conservation using classical arguments for energy conservation from [MW] or [Lu2]. \cqfd
\hspace{1cm}\\
{\bf 4 c) A space-dependent anyon problem.}\\
\\
For the space-dependent large data anyon case, the strong $L^1$-approaches available, all use a switch from time integration to 1d space integration, which in the Boltzmann case requires a 1d space setting. Under Haldane statistics and avoiding un-physical cutoffs, these strong methods are, due to the filling factors, only known to work in 2d velocity space, but that is  exactly the physically important anyon case.\\
The kernel $B(n\cdot\frac{v-v_*}{|v-v_*|},|v-v_*|)$ is assumed measurable with $0\leq B\leq B_0$. A simple cutoff is assumed; for some $\gamma, \gamma'>0$, that $B(n\cdot\frac{v-v_*}{|v-v_*|},|v-v_*|)=0$ for   $|n\cdot\frac{v-v_*}{|v-v_*|}|<\gamma'$,
for  $||n\cdot\frac{v-v_*}{|v-v_*|}|-1|<\gamma'$, and for $|v-v_*|< \gamma$, and that
$\int B(\omega,|v-v_*|))d\omega\geq c_B>0$ for $|v-v_*|\geq \gamma$.\\
The initial density $f_0(x,v)$ is assumed to be a measurable function with values in $]0,\frac{1}{\alpha }] $,  periodic in the space variable $x$,
and such that
\begin{eqnarray}
(1+|v|^2)f_0(x,v) \in L^1([ 0,1] \times \R ^2), \hspace{.2in}
\int \sup_{x\in[0,1]} f_0(x,v)dv=c_0 <\infty,\hspace{.2in}
\inf_{x\in[0,1]}f_0(x,v)>0.\hspace{.2in}
\end{eqnarray}
These conditions on $B$ are stronger, and the conditions on $f_0$ weaker than in the previous space-homogenous results.\\
With $v_1$ denoting the component of $v$ in the $x$-direction, consider for functions periodic in $x$, the initial value problem
\begin{equation}
\partial _tf(t,x,v)+v_1\partial _xf(t,x,v)= Q(f)(t,x,v),\quad f(0,x ,v )= f_0(x,v).
\end{equation}
Consider first this problem, when $f_0$ stays uniformly away from
$\frac{1}{\alpha}$ where $F(f)$ loses its Lischitz continuity. Since the gain term vanishes when $f=\frac{1}{\alpha}$ and the derivative becomes negative there, $f$ should start decreasing before reaching this value. The proof that this takes place uniformly over phase-space and approximations, requires a good control of $\int f(t,x,v)dv$ for the integration of the gain and loss parts of $Q$. That is a main topic in the proof, together with the study of a family of approximating equations with large velocity cut-off. Based on those results, and recalling the Lipshcitz continuity of $F(.)$ away from $\frac{1}{\alpha}$, contraction mapping techniques can be used to prove the well-posedness of the problem in the case when the initial value $f_0$ stays uniformly away from $\frac{1}{\alpha}$.\\
\setcounter{theorem}{1}
\begin{theorem}
{\rm [AN4]} Given $T>0$ and $\eta>0$, if $f_0$ satisfies (4.6) and $0<f_0\leq\frac{1}{\alpha}-\eta$, then there is $\eta_T$ with $0<\eta_T\leq \eta_T$ and a strong solution $f\in\mathcal{C}([0,T];L^1([0,1]\times\R^2))$ of (4.7) with $0<f\leq\frac{1}{\alpha}-\eta_T$. The solution is unique and conserves mass, first moments, and energy.
\end{theorem}
It follows that the solution $f$ exists for $t<\infty$. \\
\\
The restriction that $f_0$ stays uniformly away from $\frac{1}{\alpha}$ can be removed by a local initial layer analysis, only assuming H\"older continuity of $F$. A main ingredient in that proof is a uniform control from below of the rate of decrease of $f$ near $\frac{1}{\alpha}$.
\begin{theorem}
{\rm [AN4]} Suppose $f_0$ satisfies (4.6) and $0<f_0\leq\frac{1}{\alpha}$. Then there exists a strong solution $f\in\mathcal{C}([0,\infty[;L^1([0,1]\times\R^2))$ of (4.7) with $0<f(t,.)<\frac{1}{\alpha}$ for $t>0$. There is $t_m>0$ such that given $T>t_m$, there is $\eta_T>0$ so that $f\leq \frac{1}{\alpha}-\eta_T$  for $t_m\leq t\leq T$.
The solution is unique and is stable in the $L^1$-norm on each interval $0\leq t\leq T$. It conserves mass, first $ v$-moments, and energy.
\end{theorem}
\underline{\bf{Remarks.}} Contrary to the space-homogeneous case, here the control of $\int f(t,x,v)dv$ is non-trivial.\\
The above results seem to be new also in the fermion case ($\alpha=1$).\\
This type of approach to the classical Boltzmann equation gives in the large data $L^1$-case well-posedness for strong solutions and energy conservation.\\
The approach in the paper can also be used to obtain regularity results. \\
An entropy connected to the anyon Boltzmann equation is \\
$\int \Big(f\log f +(\frac{1}{\alpha}-f)\log (1-\alpha f)^\alpha -(\frac{1}{1-\alpha}+f)\log (1+(1-\alpha)f)^{1-\alpha}\Big)dxdp$.\\
\[\]
An open problem is the behaviour of (4.7) beyond the anyon frame, i.e. for higher dimensions under Haldane statistics. It seems likely that a close to equilibrium approach {as in the classical case,} could work with fairly general kernels $B$, but restricting the close to equilibrium $f_0$ by regularity and strong decay conditions for large velocities.
Any progress on the large data case in several space-dimensions under Haldane statistics would be quite interesting.\\
\\

\section{Conclusions; differences between quantum and classical non-linear Boltzmann theory.}
The above selection of models could be expanded to include further low temperature kinetic evolutions of collision type, such as collisions involving five quasi-particles, two colliding ones giving rise to three or conversely, cf [K]. Other examples are Grassmann algebra valued gas densities - completely anti-commuting densities e.g. for modelling the Pauli exclusion effect (cf [PR]), and pair collision terms taking account of the finite duration of a collision (cf [ES]).
But already the examples discussed, well illustrate how the situation in low temperature, non-linear quantum kinetic theory in various ways markedly differ from the classical case.\\
\\
i) { Questions from classical kinetic theory.}\\
For each type of quantum kinetic collision term, a multitude of questions from classical kinetic theory would be interesting to study. New boundary-condition dependent phenomena may appear, an example being a low-temperature gas of excitations  in a condensate between rotating cylinders having a vacuum friction type of radiation as in the Zeldovich-Starobinsky effect, cf [V].\\
ii) { New insights.}\\
In comparison with their classical counterparts, the quantum problems often require new approaches or additional ideas for their solution, as in e.g. [A3],[AN2]. \\
iii) { Qualitatively different collision operators.}\\
The type of collision operator  varies qualitatively much more in the quantum regime than in the classical one, like (1.1), (1.2), (2.1), (2.7), (3.3), (4.3),  and the examples mentioned from [K], [PR], and [ES]. The quantum influence sometimes appears as entirely new phenomena such as collision-driven spin waves in the fermionic spin matrix Boltzmann equation. The thermal de Broglie wave length may become much larger than the typical inter-particle distance, an aspect absent from the classical theory. The quantum influence leads to different filling factors, as for the NUU equation, which can stabilize as well as destabilize the collision effects. \\
iv) { Fewer collisions.}\\
There are fewer collisions close to absolute zero, where permitted energy levels usually are discrete. In the example of boson excitations plus a condensate, an excitation can interact only with the modes of the zero-point motion  that do not give away energy to it, and the domain of integration may have lower dimension than the classical context.\\
v) { Typical energy range.}\\
A quantum situation is usually not scale invariant as in the classical case, but may have a typical energy range as in Section 3. A particular type of quasi-particles/excitations exists in a particular energy interval, and extending  the energy outside this interval, may introduce collision types not observed in the experiments, and physically irrelevant for the modelling at hand. In this way restricted velocity domains, may be both physically correct and mathematically adequate in quantum situations, like the
two component, space-dependent boson example in Section 3c. Similarly, models and solutions with respect to a bounded interval of time, e.g. depending on relaxation effects, may be relevant in particular quantum situations, such as the exciton-polariton case of [DHY]. \\
vi) { Parameter range and kernel family.}\\
The questions of parameter range and kernel family are more delicate for quantum than for classical kinetic theory.
Changing a parameter mildly may completely change the mathematical aspects of the problem, as in the anyon example for $\alpha$ near $0$ and $1$. Another example where very small changes in a parameter completely changes the physical situation, is around the extremely narrow temperature-pressure domain for the phenomenologically very rich A-phase in $^3He$. The questions of parameter range and kernel family are more delicate for quantum than for classical kinetic theory. There is often no consensus about the general form of the cross section, due to a lack of proper validation analysis.\\
\hspace{1cm}\\
\hspace{1cm}\\
{\bf Bibliography}\\

\begin{itemize}

\item [] [A] T. Allemand, Derivation of a two-fluid model for a Bose gas from a quantum kinetic system. Kinetic and Related Models 2, 379-402 (2009).\\

\item [] [A1] L. Arkeryd, $L^{\infty}$-estimates for the
space-homogeneous Boltzmann equation, Jour. Stat. Phys., 31 (1983) 347-361.\\

\item [] [A2] L. Arkeryd, A kinetic equation for spin polarized Fermi systems,
Kinetic and Related Models 7, 1-8 (2014).\\

\item [] A3] L. Arkeryd, A quantum Boltzmann equation for Haldane statistics and hard forces; the space-homogeneous initial value problem, Comm. Math. Phys. 298, 573-583 (2010).\\

\item [] [AN1] L. Arkeryd, A. Nouri, Bose condensates in interaction with excitations - a kinetic model, Comm. Math. Phys. 310, 765-788 (2012).\\

\item [] [AN2] L. Arkeryd, A. Nouri, A Milne problem on the half-line with collision operator from a Bose condensate with excitations, Kinetic and Related Models 6, 671-686 (2013).

\item [] [AN3] L. Arkeryd, A. Nouri, Bose condensates in interaction with excitations - a two-component space-dependent model close to equilibrium, arXiv:1307.3012.\\

\item [] [AN4] L. Arkeryd, A. Nouri, Well-posedness of the Caucy problem for a space-dependent anyon Boltzmann equation, arXiv:1406.0265\\


\item [] [BBM] R. K. Bhaduri, R. S. Bhalero, M. V. Murthy, Haldane exclusion statistics and the Boltzmann equation, J. Stat. Phys. 82, 1659-1668 (1996).\\

\item [][BCEP]  D. Benedetto, F. Castella, R. Esposito, M. Pulvirenti, On the weak coupling limit for bosons and fermions, Math. Models Meth. Appl. Sci. 15, 1811-1843 (2005).\\

\item [] {[B]} S. N. Bose, Z. Phys. 26, 178 (1924).\\

\item [][D] J. Dolbeault, Kinetic models and quantum effects: a modified  Boltzmann equation for Fermi-Dirac particles, Arch. Rat. Mech. Anal. 127, 101-131 (1994).\\

\item [][DHY] H. Deng, H. Haug, Y. Yamamoto, Exciton-polariton Bose-Einstein condensation, Rev. Mod. Phys. 82, 1489-1537 (2010).\\

[E]  U. Eckern, Relaxation processes in a condensed Bose gas, J. Low Temp. Phys. 54, 333-359 (1984).\\

[ES] S. F. Edwards, D. Sherrington, A new method of expansion in the quantum many-body problem, Proc. Phys. Soc. 90, 3-22 (1967).\\

\item [] {[E]} A. Einstein, Sitzber. Kgl. Preuss. Akad. Wiss. 261 (1924).\\

\item [] [EH1] R. El Hajj, \'Etude Math\'ematique et Num\'erique de Mod\`eles de Transport: Application \`a la Spintronique, Th\`ese IMT, Universit\'e de Toulouse, 2008.\\

\item [] [EH2] R. El Hajj, Diffusion models for spin transport derived from the spinor Boltzmann equation, to appear in Comm. Math. Sci..\\

[ESY] L. Erd\"os, M. Salmhofer, H.-T. Yau, On the quantum Boltzmann equation, Jour. Stat. Phys. 116, 367-380 (2004).\\

[EM]  M. Escobedo, S. Mischler, On a quantum Boltzmann equation for a gas of photons, J. Math. Pures Appl. 80, 471-515 (2001).\\

[EMV] M. Escobedo, S. Mischler, M. Valle, Homogeneous Boltzmann equation in quantum relativistic kinetic theory, Electronic J. Diff. Eqns., Monograph 04 (2003).\\

[EPV] M. Escobedo, F. Pezzotti, M. Valle, Analytical approach to relaxation dynamics of condensed Bose gases, Ann. Phys. 326, 808-827 (2011).\\

[EV1] M. Escobedo, J. Vel\'azquez, Finite time blow-up and condensation for the bosonic Nordheim equation, arXiv: 1206.5410.\\

[EV2] M. Escobedo, J. Vel\'azquez, On the blow-up and condensation of supercritical solutions of the Nordheim equation for bosons, Comm. Math. Phys. 330, 331-365 (2014).\\

[FMS1] M. F\"urst, C. Mendl, H. Spohn, Matrix-valued Boltzmann equation for the Hubbard chain, Phys. Rev. E86, 031122-1-13 (2012).\\

[FMS2] M. F\"urst, C. Mendl, H. Spohn, Dynamics of the Bose-Hubbard chain for weak interactions, Phys. Rev. B89, 134311 (2014).\\

\item [] {[GNZ]} A. Griffin, T. Nikuni, E. Zaremba,  Bose-condensed gases at finite temperatures, Cambridge University Press, Cambridge 2009.\\

[H] F. D. Haldane, Fractional statistics in arbitrary dimensions: a generalization of the Pauli principle, Phys. Rev. Lett. 67, 937-940 (1991). \\

[ITG] M. Imamovic-Tomasovic, A. Griffin, Quasiparticle kinetic equation in a trapped Bose gas at low temperatures, Jour. low temp. phys. 122, 617-6555 (2001).\\

[JM] J. W. Jeon, W. J. Mullin, Kinetic equations for dilute, spin-polarized quantum systems, J. Phys. France 49, 1691-1706 (1988).\\

[JR] D. S. Jin, C. A. Regal, Fermi gas experiments, Proc. Int. School of Physics Enrico Fermi, course CLXIV, IOS Press, Amsterdam 2008.\\

[K] I. M. Khalatnikov, Theory of superfluidity (in Russian), Nauka, Moskva 1971.\\

[KK] J. Kane, L. Kadanoff, Green's functions and superfluid hydrodynamics, J. Math. Phys. 6, 1902-1912 (1965).\\

\item [] {[KD1]} T.R. Kirkpatrick, J.R. Dorfman, Transport in a dilute but condensed nonideal Bose gas: kinetic equations, J. Low Temp. Phys. 58, 301-331 (1985).\\

\item [] {[KD2]} T.R. Kirkpatrick, J.R. Dorfman, Transport coefficients in a dilute but condensed Bose gas, J. Low Temp. Phys. 58, 399-415 (1985).\\

[LM] J. M. Leinaas, J. Myrheim,  On the theory of identical particles, Nuovo Cim. B N.1, 1-23 (1977).\\

[PLL] P. L. Lions, Compactness in Boltzmann's equation via Fourier integral operators and applications I, III, Jour. Math. Kyoto Univ. 34, 391-427, 539-584 (1994).\\

[Lu1] X. Lu, A modified Boltzmann equation for Bose-Einstein particles: isotropic solutions and long time behaviour, Jour. Stat. Phys. 98, 1335-1394 (2000).\\

[Lu2] X. Lu, On isotropic distributional solutions to the Boltzmann equation for Bose-Einstein particles, Jour. Stat. Phys. 116, 1597-1649 (2004).\\

[Lu3] X. Lu, The Boltzmann equation for Bose-Einstein particles: velocity concentration and convergence to equilibrium, Jour. Stat. Phys. 119, 1027-1067 (2005).\\

[Lu4] X. Lu, The Boltzmann equation for Bose-Einstein particles: concentration in finite time, J. Stat. Phys. 150, 1138-1176 (2013).\\

[LW] X. Lu, B. Wennberg, On stability and strong convergence for the spatially homogeneous Boltzmann equation for Fermi-Dirac particles, Arch. Rat. Mech. Anal. 168, 1-34 (2003).\\

[LMS] J. Lukkarinen, Peng Mei, H. Spohn, Global well-posedness of the spatially homogeneous Hubbard-Boltzmann equation, Comm. Pure Appl. Math., May 2014 online.\\

[LS] J. Lukkarinen, H. Spohn, Not to normal order - notes  on the kinetic limit for weakly interacting quantum fields, J. Stat. Phys 134, 1133-1172 (2009).\\

[MW] S. Mischler, B. Wennberg, On the spacially homogeneous Boltzmann equation, Ann. Inst. Henri Poincar\'e 16, 467-501 (1999).\\

[N] L. W. Nordheim, On the kinetic methods in the new statistics and its applications in the electron theory of conductivity, Proc. Roy. Soc. London Ser. A 119, 689-698 (1928).\\

[No] A. Nouri, Bose-Einstein condensates at very low temperatures. A mathematical result in the isotropic case, Bull. Inst. Math. Acad. Sin. 2, 649-666 (2007).\\

[NTLCL] P. C. Nasher, G. Tastevin, M. Leduc, S. B. Crampton, F. Lalo{\"e}, Spin rotation effects and spin waves  in gaseous $^3$He, J. Physique Lett. 45, 1984, L-441. \\

[P] R. Peierls, On the kinetic theory of thermal conduction in crystals, Ann. d. Physik 3, 1055-1101 (1929).\\

\item [] {[PS]} L. Pitaevski, S. Stringari, Bose-Einstein Condensation, Clarendon Press, Oxford 2003.\\
\item [] {[PBMR]} Y. Pomeau,  M-\'E. Brachet, S. M\'etens, S. Rica,  Th\'eorie cin\'etique d'un gaz de Bose dilu\'e avec condensat, CRAS 327 S\'erie II b, 791-798 (1999).\\

[PR] S. J. Putterman, P. H. Roberts, Random waves in a classical nonlinear Grassmann field I, II, Physica A 131, 35-63 (1985).\\

[R] G. Royat, Etude de l'\'equation d'Uehling-Uhlenbeck: existence de solutions proches de Planckiennes et \'etude num\'erique, Th\`ese, Marseille 2010.\\

[ST] D. Semikoz, J. Tkachev, Condensation of bosons in the kinetic regime, Phys. Rev. D 55, 489-502 (1997).\\

[S] V. P. Silin, Introduction to the kinetic theory of gases (in Russian), Nauka Moscow, 1971. \\

%
[S1] H. Spohn, Kinetic equations for quantum many-particle systems, Modern encyclopedia of mathematical physics, Springer 2007.  \\

[S2] H. Spohn, Kinetics of the Bose-Einstein condensation, Physica D  239, 627-634 (2010). \\

[UU] E. A. Uehling, G. E. Uhlenbeck, Transport phenomena in Einstein-Bose and Fermi-Dirac gases, Phys. Rev. 43, 552-561 (1933).\\

[V] G. E. Volovik, The universe in a helium droplet, Oxford Science Publications, 2003.\\

[W] Y. S. Wu, Statistical distribution for generalized ideal gas of fractional-statistics particles, Phys. Rev. Lett. 73, 922-925 (1994).\\

\item[] {[ZNG]} E. Zaremba, T. Nikuni, A. Griffin,  Dynamics of trapped Bose gases at finite temperatures, J. Low Temp. Phys. 116, 277-345 (1999).\\

\end{itemize}

\end{document}